\documentclass[twocolumn,showpacs,preprintnumbers,amsmath,amssymb,aps, prc]{revtex4}

\begin{document}
\preprint{APS/123 QED}       
\title{\large{Deuteron photodisintegration with polarized photons at
astrophysical energies}}
\author{G. Ramachandran}
\email{gr@iiap.res.in}
\affiliation{Indian Institute of Astrophysics, Koramangala, Bangalore-560034,
India}
\author{S. P. Shilpashree}
\affiliation{Indian Institute of Astrophysics, Koramangala, Bangalore-560034,
India}
\date{\today}
\begin{abstract}
Following precise experimental studies at the Duke Free-Electron Laser
Laboratory, we discuss photodisintegration of 
deuterons with $100\%$ linearly polarized photons using a model independent 
theoretical approach taking together $M1$ and $E1$ amplitudes simultaneously.
The isoscalar $M1_s$ contribution is also taken exactly into account. From the
existing experimental measurement on doubly polarized thermal neutron capture,
it is seen that the isoscalar $M1_s$
contribution could be of the same order of magnitude
as the experimentally measured cross sections at energies relevant to Big Bang 
Nucleosynthesis (BBN).
Therefore appropriate measurements on deuteron photodisintegration 
are suggested to empirically determine the
$M1_s$ contribution at astrophysical energies.
\end{abstract}
\maketitle

The photodisintegration of the deuteron and its inverse reaction viz., n-p
fusion in the neutron
energy range of order 10 to $10^2$ keV is of considerable
interest to astrophysics. It is 
important to the nucleosynthesis scenarios \cite{1} from the Big Bang to 
stellar evolution under various conditions. The earliest estimates of the 
reaction rates by Fowler, Caughlan and Zimmerman (FCZ) \cite{2} used 
theoretical calculations \cite{3} of deuteron photodisintegration normalized to
the then available thermal neutron radiative capture cross section measurements
\cite{4}. In a comprehensive evaluation of the reaction rates and uncertainties
in 1993, Smith, Kawano and Malaney \cite{5} have pointed out : ``\textit {With a binding
energy of 2.22 MeV, deuterium is the most fragile of the primordial isotopes:
it is rapidly destroyed in stellar interiors.... Given that significant 
quantities of deuterium can only be produced during primordial nucleosynthesis,
detection of deuterium provides important evidence in favor of the big bang
model... Given the range of D/H observed in the interstellar medium, it is
difficult to directly determine a lower limit... a determination of the upper
limit is plagued by uncertainties arising from chemical evolution effects.
The ratio of the primordial abundance of deuterium to that observed today 
could be any where between 1 and 50}".

Laboratory measurements and decisive 
developments in astronomical observations go hand in hand to remove
crucial ambiguities in nuclear physics input parameters and sharpen theoretical
predictions in the astrophysical context. Although laboratory measurements
with thermal neutrons date back to 1936 by Fermi and collaborators \cite{6}
it has not been possible  for a long time to measure the cross section at
astrophysical energies due to the tendency of the neutrons to thermalize at
low energies. The first cross section measurements between 20 keV and 64 keV
have been reported in 1995 by Suzuki {\it et al.,} \cite{7} and subsequently, by
Nagai {\it et al.,} \cite{8} at 550 keV. Burles and Tytler \cite{9} measured the
deuterium abundance in high-red-shift-hydrogen clouds, where (it may be
expected that) almost none of the deuterium could have been destroyed 
subsequent to the primordial stage. However, in a re-examination of the
estimates of the uncertainties in 1999 to sharpen the predictions of big bang
nucleosynthesis (BBN), Burles, Nollett, Truran and Turner \cite{10}
have observed : ``\textit{ Our method
breaks down for the process $p+n\rightarrow d+\gamma$. This
is because of a near-complete lack of data at the energies relevant for BBN.
The approach used for this reaction is a constrained theoretical model that is
normalized to high precision thermal neutron capture cross-section
measurements}". The measured cross section of $334.2\pm 0.5 mb$ by Cox, Wynchank
and Collie \cite{11} for thermal neutrons is considered as standard.

The thermal neutron cross section has traditionally been interpretted 
in terms of a dominant isovector $M1$
amplitude for radiative capture from the initial $^1S_0$ state of the n-p 
system in the continuum. Theoretical calculations \cite{12} based on potential
models led to a $10\%$
discrepancy with the experimental measurements. Breit and Rustgi \cite{13}
proposed a polarized target-beam-test to detect the possibility of  
radiative capture from the initial $^1S_0$ state as well, which
can take place through isoscalar $M1$ and possibly also 
isoscalar $E2$ transitions. 
However, the surprising accuracy with which Riska and Brown \cite{14}
explained the $10\%$ descrepancy by including Meson Exchange Current (MEC)
contributions, set the trend for theoretical discussion in later years.
It has been noted by Nagai {\it et al.,} \cite{8} that the measured cross section is
in agreement with the theoretical calculations by Sato {\it et al.,} \cite{15}
including MEC's, isobar currents and pair currents. They have also pointed out 
that ``\textit{the theory is in good agreement with the cross section measured for
neutrons above 14 MeV, but it deviates by about 15$\%$ from the measured 
cross section of the $d(\gamma,n)p$ reaction by using the $\gamma$ ray of 
between 2.5 and 2.75 MeV \cite{16}, corresponding to neutron energies of 
550 and 1080 keV}" \cite{8}.
Experimental studies on photodisintegration
of the deuteron for photon energies from 2.62 MeV and above is well documented
 \cite{17}. The cross section at 2.62 MeV is $1.30\pm 0.029 mb$ which increases
slowly to $2.430 \pm 0.17 mb$ at 4.45 MeV and starts slowly decreasing with
energy thereafter. The disintegration process is dominantly through $E1$
transitions leading to final triplet $P$-states of the n-p system in the continuum.
Apart from the 15$\%$ discrepancy with the measured cross section noted by
Nagai {\it et al.,} \cite{8}, the measured
angular distribution and
neutron polarization at photon energy of  
2.75 MeV \cite{18} and in the range 6 to 13 MeV
 \cite{19} were found to be in disagreement with theoretical predictions 
which included the meson exchange currents. Measurements of the analyzing power
 \cite{20} in $p(\vec n, \gamma)d$ at neutron energies of 6.0 and 13.43 MeV
were consistent with \cite{19} and theoretical calculations \cite{21}
showed that meson exchange currents produce a significant change but the 
effect is to move the theoretical curve to more negative values, thus making
the discrepancy between theory and experiment more pronounced. An observable
which is sensitive to the presence of isoscalar $M1$ and $E2$ transitions from the
triplet $S$-state is the circular polarization of the emitted radiation with
initially polarized neutrons. The first measurement \cite{22} to detect the
presence of isoscalar amplitudes was not quite encouraging but a subsequent
measurement \cite{23} yielded a value $P_\gamma= -(2.29 \pm 0.9)\times 10^{-3}$.
An attempt \cite{24} to explain the large measured value by introducing a
six quark admixture in the deuteron wave function led however to a 
disagreement with the well known deuteron magnetic moment. Later calculations
 \cite{25} in the zero range approximation and the wavefunction for a Reid soft
core potential led to a theoretical prediction $P_\gamma $ of the order of 
$-1.1 \times 10^{-3}$ with an estimated accuracy of 25$\%$. The measured
value \cite{26} of $P_\gamma =-(1.5 \pm 0.3)\times 10^{-3}$ is in 
reasonable agreement with the theoretical calculation \cite{25}. The 
importance of measuring the photon polarization with initially polarized
neutrons incident on a polarized proton target has been pointed out \cite{27}.
When the initial preparation of the neutron and proton polarizations
$P(n)$ and $P(p)$ are such that they are either opposite to each
other or orthogonal to each other, the interference of the small
isoscalar amplitudes with the large isovector amplitude could substantially
contribute to the observable photon polarization.

Anticipating the experimental results of polarized thermal neutron
capture by polarized protons by M$\ddot{\rm u}$ller {\it et al.,} \cite{28},
the possibility of the initial $^3S_1$ state contributions at thermal neutron
energies was discussed using  
two different versions of effective field theory \cite{29,30}
Although the measured value of $(1.0\pm 2.5)\times 10^{-4}$ for the 
$\gamma$ anisotropy $\eta$ was not sufficiently sensitive to distinguish
between the two theoretical predictions, we may use equation (2) of
M$\ddot{\rm u}$ller {\it et al.,} \cite{28} to estimate the ratio $R$ of the triplet
to singlet capture cross sections to be $1.202 \times 10^{-3} $.
If we multiply $R$ by the well-known cross section
 \cite{11}, we 
get an estimate of 401.7 $\mu b$ for the $^3S_1$ contribution to 
the cross section at thermal neutron energies.
Quite surprisingly, this number is of the same
order as the measured cross sections for capture at astrophysical energies of 
20, 40 and 64 keV \cite{7}. In fact, it is even larger by a factor of 10 than the
measured cross section at 550 keV \cite{8}.
This raises an open question as to what could possibly be the ratio $R$ at
astrophysical energies 
relevant to BBN.
 
The influential paper of Burles, Nollett, Truran and
Turner \cite{10} has inspired several theoretical \cite{31,32} as well as 
experimental \cite{33,34} studies. Since photodisintegration of the deuteron
is well documented \cite{35} for photon energies of 2.62 MeV and above and
is known to be dominated by $E1$ transitions leading to final triplet 
$P$-states
in the n-p continuum, these studies were  motivated towards the determination
of the relative $M1$ and $E1$ contributions to the process at astrophysical
energies. The experiment \cite{33} was concerned with the measurement of the
near threshold beam analyzing power using for the first time a laser based
$\gamma$-ray source at 3.58 MeV. This was followed by measurements at seven
$\gamma$-ray energies between 2.39 and 4.05 MeV \cite{34}.
These measurements with 100$\%$ linearly polarized photons have been analyzed,
making several 
simplifying assumptions viz.,
\\ a) only $l=0,1$ partial waves were considered in the final state 
due to the low
energies involved,\\
b) of the allowed two $M1$ and four $E1$ transitions, the isoscalar $E1$ leading
to $^1P_0$ is set to zero,\\
c) the isoscalar $M1$ term leading to $^3S_1$ is neglected, using the traditional
agruments for its supression,
\\ d) the three isovector $E1$ terms were combined to form a single $P$-wave 
amplitude,\\
using the theoretical formalism \cite{36}, where $M1$ and $E1$ contributions
were calculated separately.\\

The purpose of the present paper is to study $d(\vec \gamma, n)p$ theoretically,
using a model independent formalism,
without making any simplifying
assumptions  except that only the dipole transitions are considered with $l=0,1$
 partial waves in the final state.
Since the strength of the isovector $M1_v$ amplitude which is dominant at
thermal neutron energies is known to decrease \cite{15,31,32} 
by several orders of magnitude as energies
relevant to BBN is approached and an estimate of 401.7 $\mu b$ of the 
contribution of the isoscalar $M1_s$ 
amplitude to the cross section at thermal neutron energies is seen to be of the
same order of magnitude as the measured cross sections \cite{7,8} at energies
relevant to BBN, it is not unreasonable to pay attention to the contribution
of the isoscalar $M1_s$ amplitude
at the energies of astrophysical interest.
Moreover, spin observables are generally sensitive to the interference of a 
leading amplitude with other amplitudes which are not expected to be large. It is 
therefore appropriate to study the sensitivity of the beam analyzing powers to 
the isoscalar $M1_s$ amplitude leading to final $^3S_1$
state at astrophysical energies.
\\

We choose the linearly polarized photon momentum ${\bf k}$ in c.m. frame
to be along z-axis and the linear polarization to be along
x-axis of a right handed cartesian coordinate system and the neutron
momentum ${\bf p}$ in c.m. frame to have polar coordinates $(p,\theta,\phi)$,
following \cite{33}.
If the left and right circular states of photon polarization are defined 
following Rose \cite{37} through ${\bf u}_\mu = -\mu {\bf \xi}_\mu, \mu=\pm 1$, 
the above state of
linear polarization may be represented by ${1 \over \sqrt 2} ({\bf u}_{+1} 
+{\bf u}_{-1})$. We use natural units, $\hbar = c=1$. 
The unpolarized differential cross section for the reaction
$d(\gamma, n)p$, in c.m frame at energy $E$ is given by
\begin{eqnarray}\label{dsigma0}
{d\sigma_0 \over d\Omega}&=&{1 \over 6} { E_n E_p E_d |{\bf p}| \over (2\pi E)^2 }
\sum_{\mu=-1,1} Tr({\bf T}(\mu) {\bf T}^\dagger(\mu))
\nonumber \\ 
 &=& {1 \over 6}
\sum_{\mu=-1,1} Tr[{\bf M}(\mu) {\bf M}^\dagger(\mu)],
\end{eqnarray}
where $Tr$ denotes the trace or spur and 
${\bf T}(\mu)$ denotes the on-energy-shell matrix for $d(\vec \gamma, n)p$, 
when photons are in the polarized state ${\bf u}_\mu$. The c.m. energies of the
neutron, proton and deuteron are denoted respectively by $E_n$, $E_p$ and 
$E_d$. 
Following \cite{38}, we express 
\begin{equation}
{\bf M}(\mu) = \sum_{s=0}^1 \sum_{\lambda = |s-1|}^{s+1}
(S^\lambda(s,1) \cdot {\mathcal F}^\lambda(s,\mu)),
\end{equation}
in terms of irreducible tensor operators, $S^\lambda_{\nu}(s,1)$ of rank
$\lambda$ in hadron spin space \cite{39} connecting the initial spin 1 state
of the deuteron with the final singlet and triplet states, $s=0,1$ of 
the $n-p$ system in the continuum.
Making use of the multipole expansion for ${\bf u}_\mu e^{i{\bf k}\cdot {\bf r}}$ \cite{37} 
and expressing the continuum states of the n-p system in terms of partial waves, the 
irreducible tensor amplitudes, ${\mathcal F}^{\lambda}_\nu 
(s,\mu)$ of rank $\lambda$ are given, in general, by

\begin{eqnarray}\label{irr}
{\mathcal F}^\lambda_\nu (s,\mu) &=& {1 \over 2} \sum_{L=1}^{\infty}
\sum_{l=0}^\infty \sum_{j=|l-s|}^{l+s} \sum_{I=0,1}
(i)^{L-l} \nonumber \\ & \times &  [1-(-1)^{l+s+I}] 
(-1)^{j+L-l}[L] [j]^2 [s]^{-1} \nonumber \\ & \times &
W(L1ls;j\lambda)   
F^{Ij}_{ls;L} f^\lambda_\nu(l,L,\mu),
\end{eqnarray}

where $l,I$ denote the orbital angular momentum and isospin in the final state,
$j$ denotes the conserved total angular momentum,  $L$ denotes the total angular
momentum of the photon and the shorthand notation $[L]$ stands for $\sqrt{2L+1}$.
The partial wave multipole amplitudes $F^{Ij}_{ls;L}$ depend
only on c.m. energy $E$, while the
\begin{eqnarray}
f^\lambda_\nu (l,L,\mu)&=& 4\pi \sqrt{2\pi}  (i\mu)^{\pi^+} \nonumber \\
&\times&C(l, L, \lambda ; m_l, -\mu, \nu)
Y_{lm_l}(\theta,\phi), \hspace{0.5 cm}
\end{eqnarray}
take care of the angular dependence and also the
dependence on photon polarization.
The projection operators 
\begin{equation}
\pi^{\pm} ={1 \over 2} 
\{1 \pm (-1)^{L-l}\}
\end{equation}
assume either of the values 0,1 such that, if $\pi^+ =1$ implies $\pi^- =0$ and 
vice versa. The
$F^{Ij}_{ls;L}$ denotes electric $2^L$-pole amplitudes, if $\pi^+ = 1$ and 
magnetic $2^L$-pole amplitudes, if $\pi^- = 1$. It may be noted 
that the reaction is completely characterized at any energy by the set of four
irreducible tensor amplitudes ${\mathcal F}^\lambda_\nu (s,\mu)$, given by
\eqref{irr}.
But the contributing partial wave multipole amplitudes $F^{Ij}_{ls;L}$
increase as the c.m. energy increases. 

In the region of interest to BBN, 
we may restrict ourselves to only $L=1$ and to $l=0,1$ partial waves as in \cite{33}.
Then, we clearly have two $M1$ amplitudes viz., the isovector $M1_v$ leading to 
the final $^1S_0$ state,  the isoscalar $M1_s$ leading to the final 
$^3S_1$ state  
and four $E1$ amplitudes viz., 
three isovector $E1_v^{j=0,1,2}$ leading to the final $^3P_j$ states
and an isoscalar $E1_s$ leading to the final $^1P_0$ state.
In terms of these limited number of partial wave multipole amplitudes,
the four irreducible tensor amplitudes ${\mathcal F}^\lambda_\nu(s,\mu)$ may  
explicitly be written as
\begin{eqnarray}
{\mathcal F}^1_\nu(0,\mu) &=&
-i M1_v f^1_\nu (0,1,\mu)- \sqrt 3 E1_s f^1_\nu (1,1,\mu),\\
{\mathcal F}^0_0(1,\mu)&=&{1 \over 3} E1_v(0) f^0_0(1,1,\mu), \\
{\mathcal F}^1_\nu(1,\mu) &=& -{1 \over 6} E1_v(1) f^1_\nu(1,1,\mu) 
+ i M1_s f^1_\nu (0,1,\mu),\\
{\mathcal F}^2_\nu(1,\mu) &=& {1 \over 6}E1_v(2) f^2_\nu(1,1,\mu). 
\end{eqnarray}
The $E1_v(\lambda)$ amplitudes contributing to the triplet irreducible tensor
amplitudes ${\mathcal F}^\lambda_\nu(1,\mu)$ with $\lambda =0,1,2$ are related
to the $E1_v^j$ amplitudes with $j=0,1,2$ through
\begin{equation}
\left[
    \begin{array}{c}
     E1_v(0)\\
     E1_v(1) \\
     E1_v(2)
     \end{array}
     \right ]
= \left [
\begin{array}{ccc}
1 & 3 & 5\\
2 & 3 & -5\\
2 & -3 & 1
\end{array}
\right ]
\left [
\begin{array}{c}
E1_v^{j =0}\\
     E1_v^{j =1} \\
     E1_v^{j =2}
     \end{array}
     \right ].
\end{equation}

The differential cross section relevant to \cite{33,34} for $d(\vec \gamma, n)p$
with linearly 
polarized photons is given, in c.m. frame, by 
\begin{eqnarray}
{d\sigma \over d\Omega} ={1 \over 6} Tr {\bf M M^\dagger},
\end{eqnarray}
where
\begin{equation} 
{\bf M= M}(+1)+{\bf M}(-1).
\end{equation}
Using known properties \cite{38} of the irreducible tensor operators and
standard Racah algebra, we have
\begin{equation}\label{poldiff}
{d\sigma \over d\Omega}={2\pi^2 \over 6} [a+b \sin^2\theta (1+\cos 2\phi)
-c\cos\theta],
\end{equation}
where 

\begin{eqnarray}\label{a}
a &=& \big[ 8 |M1_v|^2+ 24|M1_s|^2 + 36|E1_s|^2 +
 8 |E1_v^{j=0}|^2 \nonumber \\ &+&
18 |E1_v^{j=1}|^2 + 
26 |E1_v^{j=2}|^2 -
16 Re(E1_v^{j=0} E1_v^{j=2*}) \nonumber \\ &-&
36 Re(E1_v^{j=1} E1_v^{j=2*}) \big],
\end{eqnarray}

\begin{eqnarray}\label{b}
 b &=& \big[ 9 |E1_v^{j=1}|^2+
21 |E1_v^{j=2}|^2
+24 Re(E1_v^{j=0} E1_v^{j=2*}) \nonumber \\
&+&54 Re(E1_v^{j=1} E1_v^{j=2*})-18|E1_s|^2 \big], \hspace{1.5 mm}
\end{eqnarray}
and
\begin{equation}\label{c}
c = 4\sqrt 6 Re[(2 E1_v^{j=0} + 3 E1_v^{j=1} 
-5 E1_v^{j=2})M1_s^*\big].
\end{equation}
It is readily seen from \eqref{c} that
the third term $c \cos\theta$ in \eqref{poldiff} arises due to the
interference of the $M1_s$ amplitude with the $E1_v$ amplitudes. This term does not find
place in \cite{36}, since the calculations there have been carried out separately
for the $E1$ and $M1$ transitions. If we identify $2\pi^2 F^{IJ}_{ls;1}$ with
$32 \lambdabar^2 I_{lsb}$ of \cite{36} where $b$ denotes $j$, 
there is complete agreement 
between our expressions given by \eqref{a} and \eqref{b} 
for $a$ and $b$ and the corresponding expressions
in \cite{36}.
If it is assumed that $E1_s=0$ and
\begin{equation}\label{ass}
E1_v^{j=0} = E1_v^{j=1} = E1_v^{j=2} = E1_v,
\end{equation}
it follows that $a,b,c$ simplify to 
\begin{equation}
a= 8(|M1_v|^2 +3 |M1_s|^2) ; \quad
b=108 |E1_v|^2 ; \hspace{.5 mm}
c=0,
\end{equation}
leading to the beam analyzing power
$\Sigma (\theta)$ defined by (2) of \cite{33} which now assumes the form
\begin{equation}\label{ref33}
\Sigma(\theta) = {27 \over 2} |E1_v|^2 \sin^2\theta / D, 
\end{equation}
where the denominator 
\begin{equation}
D=|M1_v|^2 +3 |M1_s|^2+{27 \over 2} |E1_v|^2 \sin^2\theta,
\end{equation}
is proportional to the unpolarized differential cross section.
The $\Sigma (\theta)$  was determined experimentally at $\theta=150^o$ in
 \cite{33} at $\gamma-$ray energy 3.58 MeV and at $\theta=90^o$ in \cite{34}
at seven $\gamma-$ray energies between 2.39 and 4.05 MeV. The measurements
of $\Sigma (\theta)$ in \cite{33} have led to empirical estimates of 
\begin{equation}\label{X}
X = |M1|^2/|E1_v|^2 = (|M1_v|^2 +3 |M1_s|^2)/ |E1_v|^2,
\end{equation} 
if $M1_s$ is not set equal to zero. 
Under the same 
simplifying assumptions, it is interesting to note that the tensor 
target analyzing power \cite{38} is given by
\begin{equation}\label{a20}
A^2_0 = {1\over \sqrt 2}[|M1_v|^2-{3\over 2}
|M1_s|^2]/ D.
\end{equation}
Thus experimental measurements of $A^2_0$ can lead to an empirical estimate of
\begin{equation}
Y = (|M1_v|^2 -{3 \over 2}
|M1_s|^2)/ |E1_v|^2 
\end{equation}
in the energy region of astrophysical
interest. Since $X$ and $Y$ are known empirically as functions of energy, it is
possible to estimate
\begin{equation}\label{R}
R= {|M1_s|^2 \over |M1_v|^2} = {2\over 3} {(X-Y) \over (X+2Y)},
\end{equation}
to study the energy dependence of $R$ empirically in the energy region of 
interest to astrophysics.

Finally, we may point out that, when the above simplifying assumptions are not
made, the unpolarized differential 
cross section \eqref{dsigma0} itself is given by
\begin{equation}\label{diff}
{d\sigma_0 \over d\Omega} = {2\pi^2 \over 6} [a+ b \sin^2\theta -c \cos\theta],
\end{equation}
where the coefficient $c$ in third term can be determined by taking the 
difference between measurements of
${d\sigma_0 \over d\Omega}$ at two angles $\theta(\neq\pi/2$) and $\pi -\theta$.
It can also be determined in the same way from ${d\sigma \over d\Omega}$
given by \eqref{poldiff}. 
For eg., Schreiber {\it et al.,} \cite{33} have measured \eqref{poldiff} at 
$\theta = 150^o$ and for $\phi=0$ and $90^o$. Additional measurements at 
$\theta = 30^o$ for the same angles $\phi$ and at the same energy,
could easily estimate $c$ at 3.58 MeV. The measurements by Tornow 
{\it et al.,} \cite{34} at lower energies have been carried out at $\theta =
90^o$ and therefore not suitable for this purpose. It would therefore be
desirable to carryout measurements at $\theta \neq 90^o$ and at $\pi-\theta$
at lower the energy, to determine $c$.
The coefficient $b$ is readily determined by 
taking the difference between \eqref{poldiff} and \eqref{diff} 
at any angle $\theta \neq 0$ or  
$\pi$ and for any value of $\phi \neq \pi/4$. Since $b$ 
and $c$ are thus known, one can determine $a$ by measuring \eqref{poldiff} 
or even \eqref{diff}.
Thus $a, b, c$ given by \eqref{a}, \eqref{b} and \eqref{c} 
are determinable empirically without making simplifying assumptions as in
 \cite{33}. We may note from \eqref{c} that $c$ goes to zero either if
$M1_s$ is zero or if \eqref{ass} holds exactly.
On the other hand, if an empirical determination
leads to $c\neq 0$, it implies simultaneously that $M1_s \neq 0$
and the simplifying assumption \eqref{ass} is invalid.\\

Therefore an empirical 
determination of $c$ appears desirable before carrying out the more incisive
analysis of
the experimental data suggested above. 
If $c$ is found to be zero experimentally and
\eqref{ass} is assumed to be valid, the measurements of \eqref{ref33} 
along with \eqref{poldiff}, \eqref{a20}
and \eqref{diff} hold promise for the more 
incisive empirical analysis, wherein $R$ given by \eqref{R} also gets determined
as a function of energy along with \eqref{X}, where $|M1|^2$ represents
$|M1_v|^2+3|M1_s|^2$. This will lead to a better understanding of 
the photodisintegration of deuterons
at astrophysical energies of relevance for sharpening the predictions of BBN.

\begin{acknowledgements}
We are grateful to Professors B. V. Sreekantan, R. Cowsik, J. H. Sastry,
R. Srinivasan and S. S. Hassan
for the facilities provided for research at the Indian Institute
of Astrophysics.
\end{acknowledgements}

\end{document}